\documentclass[sigconf]{acmart}
\pdfoutput=1

\usepackage[T1]{fontenc}
\usepackage[utf8]{inputenc}

\usepackage{amsmath}
\usepackage{amssymb}
\usepackage{amsfonts}
\usepackage{amsthm}
\usepackage{balance}
\usepackage{graphicx}
\usepackage{tabularx}
\usepackage{color}
\usepackage[ruled,linesnumbered]{algorithm2e}
\usepackage{bm}
\usepackage{booktabs}

\newcommand*{\transp}{%
\top%
}

\newcommand{\mat}[1]{\ensuremath{\bm{#1}}}

\renewcommand{\vec}[1]{%
\bm{#1}%
}

\newcommand{\A}{%
\mat{A}%
}

\newcommand{\precond}{%
\mat{P}%
}

\newcommand{\B}{%
\ensuremath{\vec{b}}%
}

\newcommand{\x}{%
\ensuremath{\vec{x}}%
}

\newcommand{\res}{%
\ensuremath{\vec{r}}%
}

\newcommand{\dir}{%
\ensuremath{\vec{p}}%
}

\newcommand{\z}{%
\ensuremath{\vec{z}}%
}

\newcommand{\q}{%
\ensuremath{\vec{\varrho}}%
}

\newcommand{\xs}[1]{%
\ensuremath{\vec{x}^{(#1)}}%
}

\newcommand{\ress}[1]{%
\ensuremath{\vec{r}^{(#1)}}%
}

\newcommand{\dirs}[1]{%
\ensuremath{\vec{p}^{(#1)}}%
}

\newcommand{\dirps}[1]{%
\ensuremath{\vec{p'}^{(#1)}}%
}

\newcommand{\zs}[1]{%
\ensuremath{\vec{z}^{(#1)}}%
}

\newcommand{\qs}[1]{%
\ensuremath{\vec{\varrho}^{(#1)}}%
}

\newcommand{\Bs}[1]{%
\ensuremath{\vec{b}_{\Is{#1}}}%
}

\newcommand{\xss}[2]{%
\ensuremath{\vec{x}^{(#1)}_{\Is{#2}}}%
}

\newcommand{\resss}[2]{%
\ensuremath{\vec{r}^{(#1)}_{\Is{#2}}}%
}

\newcommand{\dirss}[2]{%
\ensuremath{\vec{p}^{(#1)}_{\Is{#2}}}%
}

\newcommand{\zss}[2]{%
\ensuremath{\vec{z}^{(#1)}_{\Is{#2}}}%
}

\newcommand{\I}{%
\ensuremath{I}%
}
\newcommand{\Is}[1]{%
\ensuremath{I_{#1}}%
}

\newcommand{\f}{%
\ensuremath{f}%
}

\newcommand{\nredu}{%
\ensuremath{\phi}%
}

\newcommand{\nfail}{%
\ensuremath{\psi}%
}

\newcommand{\nn}{%
\ensuremath{N}%
}

\newcommand{\problemsize}{%
\ensuremath{M}%
}

\newcommand{\queue}{%
\ensuremath{Q}%
}

\newcommand{\aspmv}{%
\texttt{ASpMV}
}

\newcommand{\period}{%
\ensuremath{T}%
}

\newcommand{\emilia}{%
\texttt{Emilia\_923}%
}
\newcommand{\audi}{%
\texttt{audikw\_1}%
}

\newcommand{\acmauthor}[7]{
\author{#1}
\orcid{#7}
\affiliation{
\institution{#2}
\department{#3}
\city{#4}
\country{#5}
}
\email{#6}
}

\acmauthor{Carlos Pachajoa}{University of Vienna}{Faculty of Computer Science}{Vienna}{Austria}{carlos.pachajoa@univie.ac.at}{0000-0002-0688-8130}

\acmauthor{Christina Pacher}{University of Vienna}{Faculty of Computer Science}{Vienna}{Austria}{christina.pacher@univie.ac.at}{}

\acmauthor{Markus Levonyak}{University of Vienna}{Faculty of Computer Science}{Vienna}{Austria}{markus.levonyak@univie.ac.at}{0000-0002-5131-6318}

\acmauthor{Wilfried N. Gansterer}{University of Vienna}{Faculty of Computer Science}{Vienna}{Austria}{wilfried.gansterer@univie.ac.at}{0000-0001-5170-1251}


\copyrightyear{2020}
\acmYear{2020}
\setcopyright{acmlicensed}
\acmConference[ICPP '20]{49th International Conference on Parallel Processing - ICPP}{August 17--20, 2020}{Edmonton, AB, Canada}
\acmPrice{15.00}
\acmDOI{10.1145/3404397.3404438}
\acmISBN{978-1-4503-8816-0/20/08}

\title{Algorithm-Based Checkpoint-Recovery for~the~Conjugate~Gradient~Method}

\begin{document}

\begin{abstract}
As computers reach exascale and beyond,
the incidence of faults will increase.
Solutions to this problem are an active research topic.
We focus on strategies to make the preconditioned conjugate gradient (PCG) solver
resilient against node failures,
specifically, the \emph{exact state reconstruction} (ESR) method,
which exploits redundancies in PCG.

Reducing the frequency at which redundant information is stored
lessens the runtime overhead.
However, after the node failure, the solver must restart from the last iteration
for which redundant information was stored,
which increases recovery overhead.
This formulation highlights the method's similarities to checkpoint-restart (CR).
Thus, this method, which we call \emph{ESR with periodic storage} (ESRP),
can be considered a form of \emph{algorithm-based checkpoint-restart}.
The state is stored implicitly,
by exploiting redundancy inherent to the algorithm,
rather than explicitly as in CR.
We also minimize the amount of data to be stored and retrieved compared to CR,
but additional computation is required to reconstruct the solver's state.
In this paper, we describe the necessary modifications to ESR to convert it into ESRP,
and perform an experimental evaluation.

We compare ESRP experimentally with previously-existing ESR and application-level in-memory CR.
Our results confirm that the overhead for ESR is reduced significantly,
both in the failure-free case,
and if node failures are introduced.
In the former case,
the overhead of ESRP is usually lower than that of CR.
However, CR is faster if node failures happen.
We claim that these differences can be alleviated by the implementation of more appropriate preconditioners.

\end{abstract}

\maketitle


\section{Introduction}
\label{sec:introduction}

In order to cover the demand for solving contemporary computational problems in reasonable times,
modern computer clusters reach unprecedented levels of parallelism.
The mean time between failures (MTBF)
in clusters formed of increasingly numerous nodes and components
will continue to drop.
In these circumstances,
there is good reason to start thinking of parallel computers as unreliable machines \cite{reed2015},
and to come up with strategies to work around this problem.

The solution of linear equations for symmetric, positive-definite (SPD) matrices
is a problem of great importance in science and engineering.
These matrices often arise from the discretization of elliptic differential equations,
describing phenomena such as heat conduction and elastic deformation of materials.
More detailed simulations require a finer grid and larger matrices.
The solution of these problems often requires computer clusters
and, if the scale of the machine is large enough,
the solver is prone to encounter faults in the computer it runs on.
To solve linear systems defined by SPD matrices,
the choice of solver often is the \emph{preconditioned conjugate gradient} (PCG) method (cf. Section \ref{sec:background:pcg}).
For very large matrices,
running PCG on large numbers of computer nodes is warranted.
Thus, it is worth to develop resilience strategies for this algorithm.

A particular mode in which faults may occur in a cluster are \emph{node failures}:
events where one or more nodes that were working on the solution of the system
become inaccessible and the information contained in them is lost.
To cope with them,
checkpoint-restart (CR) is currently the most extensively used strategy.
In CR, the state of the application is periodically stored in safe storage
(this is the checkpointing part of CR),
and in the event of a node failure,
the cluster can revert to a previously stored state and continue from there
(the restart part of CR).
Checkpointing has a runtime cost and since,
in the absence of errors,
this operation does not contribute to the results the application is producing,
it is desirable to keep the checkpointing frequency as low as possible.
Returning to the previous state, however,
incurs the runtime cost of discarding the iterations performed since the checkpoint.
Therefore, there is also pressure to checkpoint as often as possible.
Finding the optimal checkpointing period  is discussed in the literature,
and the value will depend on the incidence of errors in the machine
\cite{herault2015,young1974foa,daly2006hoe}.

The application of CR is straightforward:
The state can be the content of relevant variables selected by the programmer (application-level CR),
or simply the contents of all the memory used by it (full-memory CR).
However, there are drawbacks that the method described in this paper improves upon:
\begin{itemize}
\item It has been pointed out that CR will not scale up well on future exascale machines,
particularly in the full-memory case
\cite{cao2016,cappello2014exascale}.
\item CR strategies are generally algorithm-agnostic.
The technique consists of storing data and reverting to it,
ignoring any resilience arising from the algorithm itself,
thus potentially being suboptimal for the task at hand.
\end{itemize}

In contrast to these algorithm-agnostic approaches,
there is a category of strategies to deal with computer unreliability called \emph{algorithm-based fault-tolerance} (ABFT),
a concept first introduced in \cite{Huang1984a} where it is applied to dense matrix multiplication,
that exploits the properties of the algorithms to endow them with resilience.

In this work, we focus on the PCG method used to solve the linear system $\A{}\x{} = \B{}$,
where \A{} is an sparse SPD matrix.
PCG runs on a cluster that is vulnerable to node failures.
We expand on a previously introduced strategy: \emph{exact state reconstruction} (ESR) \cite{chen2011esr,pachajoa2018esr},
which provides ABFT resilience against node failures for PCG.
Our contributions in this work reduce its runtime overhead considerably,
particularly in the case of multiple simultaneous node failures.

\subsection{Terminology and notation}
\label{sec:introduction:terminology}

In this section, we introduce some terms that are used in this paper.

\emph{Lost nodes} are nodes that stop working in the event of a node failure:
They become inaccessible and the information contained in them is lost.
A \emph{spare node} is in standby until a node failure occurs,
at which point it replaces one of the lost nodes by reconstructing its information
and continue iterating in its place.
A spare node that, upon occurrence of a node failure,
takes the place of a lost node is called a \emph{replacement node}.
A \emph{surviving node} is still working after a node failure,
its information remains accessible, and it also continues working after the reconstruction.
In this paper, events in which a single node or multiple nodes fail simultaneously
are called single-node failure and multiple-nodes failure, respectively.

We use the notation of \cite{chen2011esr} to refer to indices of vectors and matrices.
The set of all indices is referred to as \I{}.
For a problem of size \problemsize{},
the cardinality of \I{} is \problemsize{}.
A subindex will restrict this index set to the indicated node or set of nodes:
For example, the entries corresponding to node $s$ will be denoted as \Is{s}.
If $\f{}$ is the set of nodes that fail simultaneously,
the set of indices corresponding to the lost elements is denoted as \Is{\f{}},
and we can refer to the set of indices for elements in surviving nodes as $\I{} \setminus \I_{\f{}}$.
We use index sets as subscripts to refer to entries of vectors and matrices,
for example, if we have a distributed vector $\x{}$,
we refer to the surviving vector elements as $\x{}_{\I{} \setminus \I_{\f{}}}$,
and for a matrix \A{},
the rows corresponding to the lost nodes can be written as $\A{}_{\I_{\f{}},\I}$.
We say that a node \emph{owns} an index, and thus the corresponding vector entry and matrix row,
if the index is in the set assigned to it.
In the context of the PCG method,
a superscript (as in $\xs{j}$) indicates the iteration number of a vector or scalar.

The number of nodes in the cluster is denoted with \nn{}.

The \emph{state} of an iterative solver refers to all of the solver's dynamic data,
i.e. the vectors and scalars whose values change in every iteration,
distributed among all nodes.
It does not include the static data
(system matrix, preconditioner and right-hand-side vector).
The \emph{trajectory} followed by the solver is the sequences of states that it goes through until convergence.
A given state fully defines the trajectory subsequently followed by the solver:
All future states are completely determined by the current state.

\subsection{Problem definition}
\label{sec:introduction:definition}

In this paper, we work on the solution of sparse, symmetric, positive definite linear systems
in a distributed-memory setting
using the PCG method.
The solver data is distributed, and the problem is solved in \nn{} nodes of a computer cluster.
Disjoint subsets $\I_s$ of consecutive indices are distributed among the nodes,
such that their union forms the set of all indices, $\I{}$.
Node $s$ is assigned the rows of the matrix, and entries of the vectors,
corresponding to the indices in its subset $\I_s$.
This \emph{block row distribution} is common,
and is used in prominent libraries such as PETSc~\cite{petsc-user-ref}.
The scalars used by the solver are replicated in all nodes.

We consider a situation in which the computer cluster is unreliable,
specifically, that it can suffer node failures, in which one or more nodes can fail simultaneously.
In terms of the linear solver,
these faults represent the loss of information owned by the affected nodes:
blocks of vector entries, matrix rows, and the scalars also residing in the nodes' memory.

We look for strategies to recover from these events
and still converge to the correct solution of the linear system.
In this work, the cost metric is the overall runtime for the solver to converge.
We assume the availability of sufficient memory to hold redundant information,
and of spare nodes in the cluster.

\subsection{Related work}
\label{sec:introduction:related}

Here we present literature
concerned with the recovery of linear solvers from node failures.

Concerning the checkpoint-restart approach, the currently most common way to deal with this problem,
we highlight work by Tao \cite{tao2018}.
In this thesis, the author proposes techniques for lossy compression of checkpoint data,
using prediction formulas based on spatial proximity.

In work on the solution of linear systems originating from partial differential equations (PDE),
Ltaief et al. present a strategy for resilience for parabolic PDEs against node failures~\cite{ltaief2006fta}.
Here, forward and backward-stepping strategies are described to reconstruct data in the physical domain associated to the lost node,
avoiding a more expensive checkpoint-restart approach.
The method reconstructs the iterand exactly, like the methods described in this paper,
but its application is a time-stepping solver with finite-differences discretization,
whereas ours is the conjugate gradient algorithm.
In \cite{huber2015resilience}, Huber et al. present an algorithm
for the reconstruction of the subdomain after a node failure for a multigrid solver for the Laplace equation.
The affected values are reconstructed approximately by solving a linear system local to the lost node.

Langou et al.~\cite{langou2007recovery}
work with a wider variety of iterative linear solvers.
After a node failure,
approximations to the lost entries of the iterand are found using the system matrix,
the right-hand-side vector
and the surviving data of the iterand itself
by solving a small linear system.
They can bound the new residual norm as less than the residual norm before the node failure
times a constant factor.
This method incurs no overhead in the absence of node failures.

Agullo et al. \cite{agullo2016strategies}
improve upon this strategy.
As in \cite{langou2007recovery},
they approximate the lost entries of the iterand from its surviving information,
but use least-squares minimization instead of solving a linear system.
As a result, the residual norm of the new vector will be less than or equal to
the residual norm before the node failure.

Chen \cite{chen2011esr}
introduces a way to perform exact state reconstruction (ESR) for multiple iterative methods,
including PCG.
They present a strategy to exploit the sparse matrix-vector product (SpMV) product to store redundant information for the input vector,
so that the full state of the vector can be reconstructed.
In \cite{pachajoa2018esr}, Pachajoa et al. extend the algorithm in \cite{chen2011esr}
to combine ESR with differently formulated preconditioners (the preconditioner itself, its inverse or a split preconditioner),
and also compare ESR to the linear interpolation algorithm from \cite{langou2007recovery}.
In \cite{pachajoa2019esr}, Pachajoa et al.
extend the ESR approach by describing how to operate in the event of multiple, simultaneous node failures.

In \cite{levonyak2020ppcg}, Levonyak et al. extend the concept of ESR
to the pipelined PCG algorithm,
while maintaining its communication-hiding properties.

The work mentioned so far supposes the availability of spare nodes.
In \cite{hori2020}, Hori et al. propose strategies for the allocation of these spare nodes, and the replacement of lost nodes, when runtime performance is of consideration.

Pachajoa et al. \cite{pachajoa2019nospare} introduce an ESR method which does not require spare nodes as replacements for failed nodes,
but can reconstruct the lost information and continue on the surviving nodes.

\subsection{Contributions of this paper}

As the main contribution of this paper,
we frame ESR for PCG as an instance of what we call \mbox{\emph{algorithm-based~checkpoint-restart}},
and describe how it can be restructured to enable decreased state-storage frequencies.
We experimentally show that this approach reduces the runtime overhead in the absence of node failures.
This is particularly beneficial in scenarios with multiple node failures,
where the additional communication needs increase the overhead most drastically and,
consequently,
for which the runtime overhead reduction is greatest.

\medskip

The rest of this paper is structured in the following manner:
In Section \ref{sec:background},
we describe the exact state reconstruction approach applied to PCG in more detail.
In Section \ref{sec:periodic},
we reframe ESR, as presented in \cite{pachajoa2019esr},
as a CR-like method,
for which the state-storage interval can be optimized,
and which offers reduced runtime overheads.
In Section \ref{sec:setup},
we describe the framework we use to obtain our experimental results.
In Section \ref{sec:experiments},
we present our experimental results,
highlighting favorable scenarios for new methods.
Finally, Section \ref{sec:conclusions} concludes the paper and presents our perspectives on future work.

\section{Algorithmic background}
\label{sec:background}

In this section, we introduce the PCG method,
describe the way we exploit the inherent data redundancies in the sparse matrix vector-product,
and explain the exact state reconstruction method.

\subsection{Preconditioned conjugate gradient}
\label{sec:background:pcg}

PCG is a linear solver for the system $\A{}\x{} = \B{}$,
where \A{} is an SPD matrix.
The method is applied in conjunction with a preconditioner,
which reduces the number of iterations until convergence
at the price of the application of the preconditioner in every iteration.

The variables used in PCG are the following:
\x{} is the iterand vector,
containing the current approximation to the solution.
\precond{} is the preconditioner, here representing its action as a linear operator.
The residual vector is represented with \res{},
and the preconditioned residual vector with \z{}.
The search direction vector, \dir{},
determines the direction in which the iterand is modified in every iteration.
$\beta$ is a scalar used for the conjugation of the search directions,
and $\alpha$ is a scalar determining the length of the step to be taken along the search direction towards the solution.
PCG is presented in Alg.~\ref{alg:pcg}.

In exact arithmetic,
supposing a naive selection of the initial guess,
solving an SPD linear system of size \problemsize{}
will take the conjugate gradient method \problemsize{} iterations to reach the solution.
If the solver is restarted from the iterand at some point before convergence,
reinitializing the search directions,
reaching the solution might require performing \problemsize{} additional iterations from that point,
thus wasting the work already performed.
In \cite{pachajoa2017cgmg},
it is shown that this effect is also observed in floating-point arithmetic.
This observation is the motivation for the \emph{exact state reconstruction} (ESR) method (cf. Section \ref{sec:background:esr}).

\begin{algorithm}
\fontsize{9}{14}\selectfont
\(\ress{0} \coloneqq \B{} - \A{} \xs{0}, \zs{0} \coloneqq \precond{} \ress{0},
\dirs{0} \coloneqq \zs{0}\)\;

\For{\(j = 0, 1, \dots, \text{ until convergence}\)}{
\(\alpha^{(j)} \coloneqq \res{}^{(j) \transp} \zs{j} / \dir{}^{(j) \transp} \A{} \dirs{j}\)\; \label{alg:pcg:alpha}
\(\xs{j+1} \coloneqq \xs{j} + \alpha^{(j)} \dirs{j}\)\; \label{alg:pcg:x}
\(\ress{j+1} \coloneqq \ress{j} - \alpha^{(j)} \A{} \dirs{j}\)\; \label{alg:pcg:r}
\(\zs{j+1} \coloneqq \precond{} \ress{j+1}\)\; \label{alg:pcg:z}
\(\beta^{(j)} \coloneqq \res{}^{(j+1) \transp} \zs{j+1} / \res{}^{(j) \transp} \zs{j}\)\; \label{alg:pcg:beta}
\(\dirs{j+1} \coloneqq \zs{j+1} + \beta^{(j)} \dirs{j}\)\; \label{alg:pcg:p}
}
\caption{Preconditioned conjugate gradient (PCG) method
  \cite[Alg.~9.1]{saad2003iterative}}
\label{alg:pcg}
\end{algorithm}

\subsection{Augmented sparse matrix-vector product}
\label{sec:background:aspmv}

In order to reconstruct the entirety of the state of the PCG solver as it was before a node failure,
enough redundant information has to be available.
That is, there has to be redundancy that we can exploit.

The SpMV already provides some redundancy:
In order to compute the product of matrix \A{} and vector \dir{},
entries of \dir{} must be transmitted from their owner node
to other nodes in the cluster,
thus already creating copies of some entries in other locations.
However, in order for the SpMV to provide full redundancy for the input vector,
the matrix must fulfill the following condition \cite{chen2011esr}:
For every node $s$, every column of the submatrix $\A{}_{\I{} \setminus \I{}_s, \I{}_s}$
contains at least one non-zero entry.
For a matrix with this property,
every entry of the vector is communicated
from its owner to at least one other node.
Most matrices do not fulfill this condition.
Furthermore, this would only guarantee sufficient redundancy to recover from the failure of a single node.

To achieve the required redundancy,
we use the extensions to the SpMV introduced in \cite{chen2011esr,pachajoa2019esr}
and name the concept \emph{augmented sparse matrix vector product} (ASpMV).
Entries that would not have been sent to any node with the ordinary SpMV are transferred to a neighbor anyway.
With our chosen strategy, node $s$ sends a given entry to node $(s+1)\,mod\,N$,
if this entry is not already being sent to some other node as part of the regular SpMV.

The exact communication overhead depends on the sparsity pattern of the matrix.
In general, denser matrices will have lower overheads for ASpMV,
since more information has to be sent anyway to compute the product.
With the nodes sending information to their neighbors,
it is convenient if the matrix is banded, with most of its entries close to the diagonal.
That way, the amount of information that ASpMV has to send additionally to the neighbors is minimized.

\subsubsection{Redundancy against multiple-nodes failures}
\label{sec:background:aspmv:multiple}

ASpMV can also guarantee the presence of several redundant copies of each input vector element.
This is necessary in order to provide resilience against multiple-nodes failures.
In order to describe this extension,
we make use of the notation of \cite[\S 3 and \S 4]{pachajoa2019esr}:
Let \nredu{} denote the target number of times each entry of the input vector must be replicated in the cluster
(i.e. the number of simultaneous node failures that should be supported).
Before, we defined $I_s$ as the set of all indices owned by node $s$.
Let $I_{s,l}$, with $l\in\{1..\nn{}\}$, be the subset of $I_s$
with indices of the input vector $\dir{}$
corresponding to entries that must be sent to node $l$ for the computation of the product $\A{} \dir{}$.
Node $s$ does not send data to itself during this operation, so we define $I_{s,s}\coloneqq \emptyset$.
Furthermore, let $d_{s,k}$ denote the designated destination nodes for resilient copies of the vector elements of node $s$,
with $k\in\{1..\nredu{}\}$.
In this work, we select $d_{s,k}$ to be the \nredu{} nearest neighbors of node $s$.
This can be achieved with the following strategy:

\begin{equation}
\label{eq:multiple:rec}
d_{s,k} \coloneqq \left\{
\begin{array}{ll}
\left( s + \left\lceil \frac{k}{2} \right\rceil \right) \bmod \nn{}, &
  \text{if}~k~\text{odd} \\
\left( s - \frac{k}{2} \right) \bmod \nn{}, & \text{if}~k~\text{even}
\end{array}
\right.
\end{equation}

For index $i$,
which belongs to node $s$,
we define its multiplicity, $m(i)$,
as the number of subsets $I_{s,l}$, with $l\in\{1..\nn{}\}$, where $i$ is present.
That is, $m(i)$ is the number of nodes that the $i^{th}$ vector entry must be sent to
in order to compute $\A{} \dir{}$.
Additionally, we define $g(i)$ as the number of the subsets $I_{s,d_{s,k}},\,k\in\{1..\nredu{}\}$ in which $i$ appears,
that is, how many of the nodes $d_{s,k}$ already need the $i^{th}$ entry to compute $\A{} \dir{}$.

We can now describe the set $R_{s,k}^c$ of indices of entries, owned by node $s$, to be sent to $d_{s,k}$, in addition to the ones required to compute the product $\A{} \dir{}$:

\begin{equation*}
R_{s,k}^c \coloneq \left\{ i \in I_s \, | \, i \notin I_{s,d_{s,k}} \;\text{  and  }\; m(i) - g(i) < \nredu{} - k \right\}.
\end{equation*}

That is, the $i^{th}$ entry, if owned by node $s$, will be sent to $d_{s,k}$ if
(1) it is not already being sent there and if
(2) as we traverse the designated destination nodes by increasing $k$,
the target number of copies for this entry has not been met yet.
The approach for a single-node failure described earlier in this section is
the same as the approach for a multiple-nodes failure,
with \nredu{} set to $1$.

After the ASpMV is complete,
each entry of the vector will have been communicated by its owner to at least \nredu{} nodes,
thus creating $\nredu{} + 1$ copies (one of them in the owner).
If up to \nredu{} nodes fail simultaneously,
each entry will have survived in at least one node
and can afterwards be transferred to a replacement node.

With this method, we send entries that are not necessary for the computation of the
matrix-vector product.
The communication of this additional information will cause iterations to take longer,
and thus lead to an increased runtime until convergence.
The exact overhead depends on factors such as the sparsity pattern of the matrix
and the network topology of the cluster.
Optimization of our strategies taking these factors into consideration is beyond the scope of this paper.
Research in this direction is ongoing work.

\subsubsection{Redundant copies}
\label{sec:background:aspmv:redundant}

After the ASpMV is executed,
the redundant information of the search direction \dirs{j} is not explicitly available.
This means, even though the information of the copies is present and spread in the cluster,
after a node failure, the data must be gathered in a replacement node before we can work with \dirs{j} again.
We introduce the concept of a \emph{redundant copy},
designated with a prime symbol ($'$),
as in \dirps{j},
to abstractly represent the redundant vector data in the cluster,
in whatever storage scheme the framework utilizes.
We do not specify the number of copies per vector entry that a redundant copy represents,
since we do not need this information for further descriptions.
The concept of redundant copies will be used to explain our algorithms in Section~\ref{sec:periodic}.

\subsection{Exact state reconstruction for PCG}
\label{sec:background:esr}

The PCG algorithm performs a matrix-vector product in every iteration
(Line \ref{alg:pcg:alpha} of Alg. \ref{alg:pcg}):
The system matrix \A{} is multiplied with the search direction vector \dir{},
thus potentially providing redundancy for the latter.
In \cite{pachajoa2019esr},
the authors explain how to reconstruct
the state of the solver as it was prior to the node failure,
save for small perturbations resulting from floating-point arithmetic.
With the two latest search directions,
and after retrieving the scalar $\beta$
from one of the surviving nodes,
it is possible to move backwards
from Line \ref{alg:pcg:p} of Alg. \ref{alg:pcg}
and reconstruct every vector involved in the computation.
The reconstruction procedure,
run on replacement nodes,
is presented in Alg.~\ref{alg:esr}.
Note that the reconstruction procedure assumes that the static solver data
(system matrix, preconditioner and right-hand-side vector)
can be retrieved from safe storage.

\begin{algorithm}
\fontsize{9}{14}\selectfont

Retrieve the static data \(\A{}_{\I{}_{\f{}}, \I{}}\), \(\precond{}_{\I{}_{\f{}}, \I{}}\),
  and \Bs{\f{}}\;
Gather \(\ress{j}_{\I{} \setminus \I{}_{\f{}}}\) and
  \(\xs{j}_{\I{} \setminus \I{}_{\f{}}}\)\;
Retrieve the redundant copies of \(\beta^{(j-1)}\), \dirss{j-1}{\f{}}, and
  \dirss{j}{\f{}}\; \label{alg:esr:pcg:2:retrieve}
Compute \(\zss{j}{\f{}} \coloneqq \dirss{j}{\f{}} - \beta^{(j-1)} \dirss{j-1}{\f{}}\)\;
  \label{alg:esr:z}
Compute \(\vec{v} \coloneqq \zss{j}{\f{}} -
  \precond{}_{\I{}_{\f{}}, \I{} \setminus \I{}_{\f{}}} \res{}_{\I{} \setminus \I{}_{\f{}}}^{(j)}\)\;
  \label{alg:esr:pc_rhs}
Solve \(\precond{}_{\I{}_{\f{}}, \I{}_{\f{}}} \resss{j}{\f{}} = \vec{v}\) for \resss{j}{\f{}}\;
  \label{alg:esr:pc_solve}
Compute \(\vec{w} \coloneqq \Bs{\f{}} - \resss{j}{\f{}} -
  \A{}_{\I{}_{\f{}}, \I{} \setminus \I{}_{\f{}}} \xs{j}_{\I{} \setminus \I{}_{\f{}}}\)\;
  \label{alg:esr:rhs}
Solve \(\A{}_{\I{}_{\f{}}, \I{}_{\f{}}} \xss{j}{\f{}} = \vec{w}\) for \xss{j}{\f{}}\;
  \label{alg:esr:pcg:2:solve}

\caption{ESR reconstruction phase for the PCG method on the replacement nodes \cite[Alg.~2]{pachajoa2018esr}}
\label{alg:esr}
\end{algorithm}

With the state of the solver as it was before the node failure,
it is possible to reach convergence following the same trajectory as an undisturbed solver.
As illustrated in \cite{pachajoa2019esr},
this method produces very low overheads,
particularly if it protects only against single-node failures.

\section{ESR with periodic storage}
\label{sec:periodic}

In this section,
we extend ESR introduced in Section \ref{sec:background:esr}
to perform iterations with ASpMV with a reduced frequency,
that is, not in every iteration, but two consecutive times every \(T\) iterations.
We call \(T\) the \emph{checkpointing interval} to keep with CR terminology,
and we refer to the set of two iterations in which redundant information is stored
as the \emph{storage stage}.
In the event of a node failure,
the solver will return to the last time the search directions for two
successive iterations were stored redundantly via ASpMV.
It is then possible to reconstruct the state for the last of those iterations.
We call the new approach \emph{Exact state reconstruction with periodic storage} (ESRP),
and contrast it with ESR
which stores data in every iteration.

To describe ESRP,
we introduce the concept of a \emph{queue},
where the solver stores redundant copies
(cf. Section \ref{sec:background:aspmv:redundant}).
In the ESR algorithm,
this queue has space for two positions:
Every iteration, ASpMV will push a new redundant copy into the queue,
and the oldest copy will be released.
This queue thus contains the redundant copies of search directions
for two successive iterations.

We now examine the same procedure in the case of ESRP.
Suppose that the last redundant copies held in the queue are \dirps{j} and \dirps{j+1},
that we performed some additional iterations using regular SpMV afterwards,
and that then a node failure occurs.
The search directions in the queue
could be used to reconstruct the state for iteration $j+1$.
However, it is possible that the node failure takes place after only one of the two iterations of a storage stage have been completed.
Suppose that the solver has reached a storage stage at some iteration $j$,
and the first call to \texttt{ASpMV} is performed.
The redundant copy \dirps{j} is then pushed to the queue.
If a node failure happens at this point in time,
before the redundant copy \dirps{j+1} is created,
the vector \dirs{j} that we can retrieve is not sufficient to perform the reconstruction shown in Alg. \ref{alg:esr},
since we would additionally need \dirs{j+1}.
For this reason, it is necessary to have a queue of not two, but three redundant copies of search directions,
such that, if this happens,
the queue still contains entries of two successive search directions from a redundant storage period before.

In addition to the redundant copies created during the ASpMV,
the solver needs to duplicate some local data at each node during the storage stage.
As can be seen in Line \ref{alg:esr:z} of Alg. \ref{alg:esr},
to reconstruct the vector \zs{j},
and subsequently the state of the solver for iteration $j$,
the value of \(\beta^{(j-1)}\) is needed.
However, depending on when the node failure occurs,
the scalar \(\beta\) may have changed since the last storage stage.
It is therefore necessary to create a duplicate of the value of \(\beta^{(j-1)}\).
Similarly, the local entries of the residual \res{}, the preconditioned residual \z{},
the iterand \x{} and the search direction \dir{}
at iteration \(j\) must also be duplicated in all nodes,
so that they can be used for the reconstruction process
and so that the surviving nodes can reset their own parts of the solver state
to match the state that is reconstructed at the replacement nodes.
Entries of the vector \zs{j} corresponding to surviving nodes are not used during the reconstruction,
and could also be recomputed from \ress{j} once the latter has been reconstructed.
However, our solver stores a local copy instead of performing this operation.
We mark these duplicate values with an asterisk:
\(\beta*\) is a scalar, and  \(\res{} *\), \(\dir{}*\), \(\z{}*\) and \(\x{} *\) are distributed vectors.
There is no need to store the scalar \(\alpha\).
It is not used during the reconstruction
and it will be computed in Line \ref{alg:pcg_periodic:alpha} of Alg. \ref{alg:pcg_periodic}
when the solver continues iterating.
Since these copies are created locally by each node,
they do not introduce any additional communication between nodes,
and the runtime overhead they cause is therefore negligible.
Note that since \(\res{} *\), \(\dir{}*\), \(\z{}*\) and \(\x{} *\) are created by each node copying its own data,
if a node fails, the copies contained in it are also lost.
Therefore, these copies, by themselves,
obviously cannot be used to reconstruct the state.

An example of the procedure follows:
The solver has a queue \queue{} with three positions to hold redundant copies.
Initially, the queue is empty:
\(Q \coloneqq [\_,\_,\_]\).
After \period{} iterations,
\aspmv{} will be called for the first time
and will push the first redundant copy,
thus \queue{} contains \([\_,\_,\dirps{T}]\).
The solver will also create a copy of the value of \(\beta^{(T)}\), $\beta*$, on every node.
At this point, it is not possible to recover from a node failure using ESRP.
After another iteration,
\aspmv{} pushes another redundant copy,
\queue{} becomes \([\_,\dirps{T}, \dirps{T+1}]\),
and copies of the vectors \ress{T+1}, \zs{T+1}, \xs{T+1} and \dirs{T+1} are made,
respectively designated $\res{}*$, $\z{}*$, $\x{}*$ and $\dir{}*$.
This information can now be used to reconstruct the state for iteration \(T+1\),
and the storage stage ends.
From there, we continue iterating using regular \texttt{SpMV}.

\begin{figure*}
\begin{center}
\includegraphics{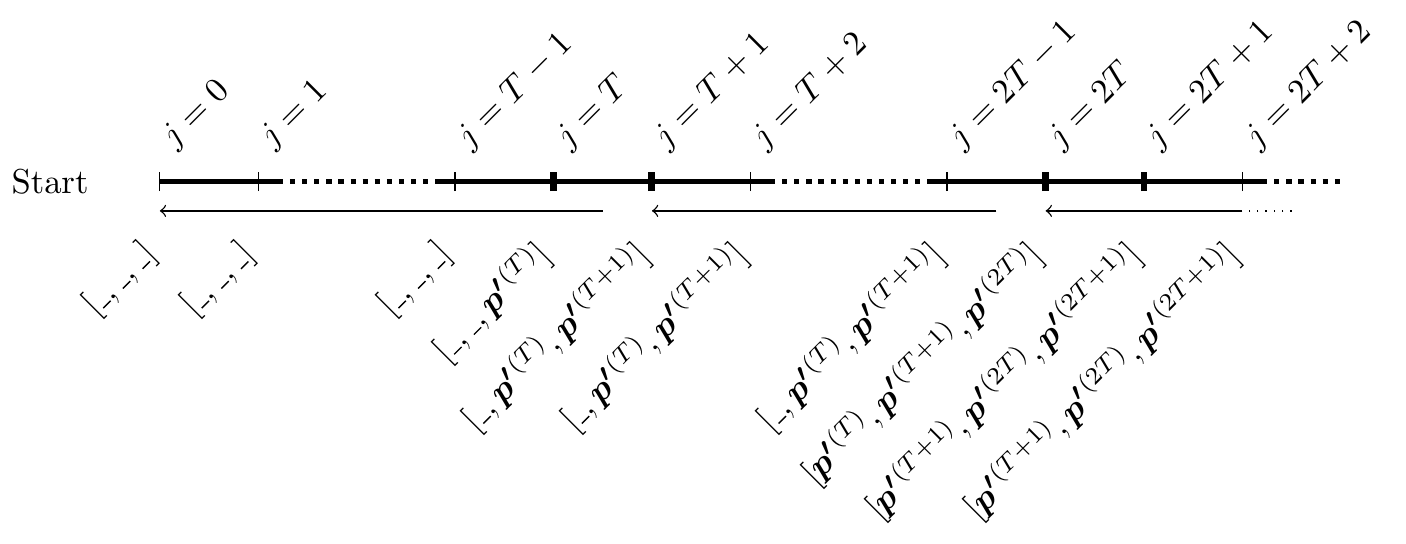}
\end{center}
\caption{
State of the redundancy queue for the search directions during the solution process.
The lists below the line represent the state of the search direction queue,
and which search directions are stored redundantly somewhere in the cluster.
The thin arrows running leftwards show how far the solver has to revert in the event of a node failure.
}
\label{fig:storage_chain}
\end{figure*}

{
\sloppy
When the next storage stage is reached,
after an additional \(T\) iterations,
we use \texttt{ASpMV} again,
and \queue{} becomes
\mbox{\([\dirps{T}, \dirps{T+1},\dirps{2T}]\)},
and again, a copy of \(\beta^{(2T)}\) is created.
The newest entry of \queue{} cannot be used in conjunction with the previous ones
to reconstruct the state of the solver.
In the event of a node failure at this point,
the state would be recovered for iteration \(T+1\).
At this time, we still need the value of \(\beta^{(T)}\) to reconstruct the state,
so we may not overwrite \(\beta*\) yet,
and will overwrite it in the next iteration instead.
After an additional successful iteration, the queue contains \([\dirps{T+1},\dirps{2T},\dirps{2T+1}]\),
and copies for \ress{2T+1}, \zs{2T+1}, \xs{2T+1} and \dirs{2T+1} are created.
From that point on, it is possible to reconstruct for iteration \(2T+1\),
which concludes the second storage stage.
The process is presented in Alg.~\ref{alg:pcg_periodic}
and graphically represented in Fig. \ref{fig:storage_chain}.
For the description of the algorithms in Section \ref{sec:periodic},
we refer to the ordinary SpMV as \(\q{} \coloneqq \text{\texttt{SpMV}}(\A{}, \dir{})\),
a function that takes matrix \A{} and vector \dir{} as inputs and returns their product, \q{}.
The augmented variant is represented as \(\q{} \coloneqq \text{\texttt{ASpMV}}(\A{},\dir{},\nredu{},\queue{})\),
where the function additionally takes the number of desired redundant copies \nredu{}, and the queue \queue{}
where it will push a new \emph{redundant copy} for \dir{}.
}

\begin{algorithm}
\fontsize{9}{12}\selectfont

\(\ress{0} \coloneqq \B{} - \A{} \xs{0}, \zs{0} \coloneqq \precond{} \ress{0},
    \dirs{0} \coloneqq \zs{0},j \coloneqq 0,\) \par
    \( \queue{} \coloneqq [\_,\_,\_] \)\;

\Repeat{\(\|\res{}\|_2 / \|\B{}\|_2 < rtol\)}{
  \uIf {\( j\;\text{mod}\;T = 0\) \enspace and \enspace \(j > 2\)}{
    \(\qs{j} \coloneqq \text{\texttt{ASpMV}}(\A{},\dirs{j},\nredu{}, \queue{})\)\;
    \(\beta** = \beta^{(j)}\)\;
  }
  \uElseIf {\enspace \((j-1)\;\text{mod}\;T = 0\)\enspace and \enspace \(j > 2\)}{
    \(\qs{j} \coloneqq \text{\texttt{ASpMV}}(\A{},\dirs{j},\nredu{}, \queue{})\)\;
    \(\x{}* = \xs{j}\),\, \(\res{}* = \ress{j}\),\, \(\z{}* = \zs{j}\),\, \(\dir{}* = \dirs{j}\)\;
    \(\beta* = \beta**\)\;
  }
  \uElse{
    \(\qs{j} \coloneqq \text{\texttt{SpMV}}(\A{},\dirs{j})\)\;
  }
  \(\alpha^{(j)} \coloneqq \res{}^{(j) \transp} \zs{j} /
    \dir{}^{(j) \transp} \qs{j}\)\;
    \label{alg:pcg_periodic:alpha}
  \(\xs{j+1} \coloneqq \xs{j} + \alpha^{(j)} \dirs{j}\)\;
    \label{alg:pcg_periodic:x}
  \(\ress{j+1} \coloneqq \ress{j} - \alpha^{(j)} \qs{j}\)\;
    \label{alg:pcg_periodic:r}
  \(\zs{j+1} \coloneqq \precond{} \ress{j+1}\)\;
    \label{alg:pcg_periodic:z}
  \(\beta^{(j)} \coloneqq \res{}^{(j+1) \transp} \zs{j+1} /
    \res{}^{(j) \transp} \zs{j}\)\;
    \label{alg:pcg_periodic:beta}
  \(\dirs{j+1} \coloneqq \zs{j+1} + \beta^{(j)} \dirs{j}\)\;
    \label{alg:pcg_periodic:p}

  \(j \coloneqq j+1\)\;
}

\caption{Preconditioned conjugate gradient (PCG) method
with periodic redundant storage (for ESRP)}
\label{alg:pcg_periodic}
\end{algorithm}

From Fig. \ref{fig:storage_chain} we can see which checkpointing intervals make sense for ESRP.
For $T>2$, we proceed as explained above.
For $T=2$, it no longer makes sense to use this approach,
since redundant copies are created in every iteration.
In this case, it is better to use ESR (Section \ref{sec:background:esr}).
For $T=1$, we can no longer talk about creating two successive copies.
Again, this corresponds to regular ESR.

\subsection{ESR/ESRP and checkpointing}

The addition of a checkpointing interval to ESRP highlights its similarity to CR approaches.
In both cases, we store the state of the solver after every checkpointing period,
either explicitly, in the case of CR,
or implicitly, by exploiting redundancy provided by the algorithm itself,
in the case of ESRP.
We claim, therefore, that ESRP is an \emph{algorithm-based checkpoint-restart} strategy.

As with CR, this method introduces a trade-off between the runtime overhead,
which decreases with increasing $T$ since we create redundant copies less frequently,
and the cost of discarding the iterations performed since the last storage stage was reached.

\subsubsection*{In-memory checkpoint-restart}

In our experiments in Section \ref{sec:experiments} we compare the runtime of our ESRP approach
to an in-memory buddy checkpoint-restart strategy (IMCR),
which will now be described in detail.

Similar to the description above,
we assume a checkpoint interval of \(T\):
Once every \(T\) iterations,
each node will create a checkpoint by sending a complete copy of the local parts of all vectors it owns
to a neighboring node (the ``buddy node'').
In the event of a node failure,
the replacement node will then simply retrieve its local vector parts from the buddy.
As in the case of ESR,
we assume that the static solver data can be retrieved from safe storage
and does not need to be stored during the checkpointing.
To extend this approach to support multiple node failures,
it is sufficient to send the checkpointing data to multiple buddies.
This is a form of \emph{algorithm-based fault tolerance},
since the checkpointing and recovery strategies are specifically tailored to the PCG solver.

There are further similarities between this checkpointing strategy and ESR/ESRP:
For example, the data that can be retrieved from a checkpointing buddy
is the same data that is reconstructed during the recovery phase of ESR,
and the strategy for choosing the buddies is the same as the strategy for
determining the destinations of redundant elements in the augmented sparse matrix-vector product,
as defined by Eq.~\ref{eq:multiple:rec} in Section \ref{sec:background:aspmv:multiple}.
An important difference, however,
is that ESR mainly adds on to existing communication,
while the checkpointing strategy introduces a completely new round of communication in each storage iteration.

\section{Implementation}
\label{sec:setup}

We implement our algorithms using our own framework,
written in C.
In this way we achieve the highest flexibility for our purpose.
Elementary linear algebra functionality is provided by GSL~\cite{gsl-manual},
but parallelization of linear algebra operations,
in particular of the matrix-vector product,
is in-house code.
Communication between nodes is realized with MPI.
The framework is modularly structured,
such that different strategies to achieve data redundancy,
and to perform reconstruction and recovery,
can be used.
In particular, we can simulate ESRP, as well as in-memory CR.

We simulate one node failure event for each run of PCG.
The ranks of the affected nodes and the iteration at which the failure should occur
are passed as parameters to our framework.
Once the marked iteration is reached,
the nodes set to fail zero-out all their vector entries,
as well as the scalars they contain,
thus simulating the loss of all of their dynamic data
(their components of vectors \x{}, \res{}, \z{}, \dir{},
as well as the scalars $\beta$ and $\alpha$).
This is also the initial state of a replacement node,
which starts without knowledge of the state of the node it is replacing.
For the sake of ease of implementation,
the set of nodes simulating a node failure will also act as the replacements.
After a simulated node failure, the replacement nodes start the recovery process,
collecting information from their neighbors and reconstructing the lost data.

In a real-world scenario,
the replacement nodes would also have to reload the rows of the system matrix and the preconditioner,
and the entries of the right-hand-side vector that they own;
however, we have decided not to include this step in the measurement of the runtime overheads during our experiments.
There are two main reasons for this decision.
Firstly, this overhead depends too strongly on the individual use case
for us to be able to make any generalized statements about it:
Not only is it influenced by the matrix size and file system properties,
but matrices might be stored in different file formats (e.g. plain text or binary),
or the solver might be working with a matrixless representation altogether.
This changes the loading time considerably.
Secondly, the reloading step is the same for both the ESR and CR versions we investigate.
Therefore, we would not gain any valuable information
about differences in the behaviour of these strategies
from examining the time required for the reloading of static data.

\subsection*{Beyond node-failure simulation}

Since we only simulate node failures,
our framework does not capture all of the events that would take place in the case of a real
incident, nor all of the steps that are necessary to recover from one.
We assume that, in a realistic application,
there would be some middleware available to take care of these additional tasks.
At any rate,
we can describe the operations necessary to perform the recovery using ESR and CR
that are not modeled in our framework.

The first of these tasks is detecting a node failure:
A prerequisite for recovery for both ESR and CR
is that there is a mechanism in place that will notice if one of the nodes becomes unresponsive.
It is reasonable to assume that this cost would be the roughly the same for both approaches in a framework that is well-optimized for this purpose.

A second task currently not modeled is determining which node has failed.
The surviving nodes need this information to decide what data must be transferred to the replacement node,
and the replacement node needs it to know which rows of the matrix, the preconditioner, and which entries of the right-hand-side vector to load.

A third task would be to setup the cluster to continue working.
For ESR as as well as the CR strategy described in this paper,
this involves providing a replacement node to take the place of the lost node
and setting up a new communicator to continue iterating.
In the case of ESR, it is also possible to proceed without a replacement node
(cf.~\cite{pachajoa2019nospare}),
but this is beyond the scope of this paper.
More generally speaking,
depending on how exactly CR is implemented,
it could make use of spare nodes,
enabling the application to keep using most of the nodes already allocated to it,
but making it necessary to identify the identity of the lost node
just as in the case of ESR;
or the whole application could be restarted on newly-allocated nodes,
although this is likely to be more costly that identifying the lost nodes,
particularly at greater scales \cite{chakraborty2015pmi,teranishi2014ulfm,hori2020}.

All in all,
we expect the costs of the events that our framework is not modeling to be comparable between ESR and CR.

Although presently not in the MPI standard,
there is ongoing work on tools to deal with node failures.
The \emph{User-Level Fault Mitigation} (ULFM) library \cite{bland2013ulfm,MPIF2017a}
offers functions for detection of node failures and identification of the affected nodes.
In conjunction with standard MPI,
it is possible to create a new communicator on which the solver can continue working.

\section{Experiments}
\label{sec:experiments}

\subsubsection*{Experimental setup}

Our experiments are run on the VSC3 machine of the Vienna Scientific Cluster.
We use 128 nodes,
with one process per node.
(One process is sufficient to examine the overheads for resilience,
since the redundant data has to be sent to different nodes in any case.)
This machine has a fat-tree topology.

We use the following libraries:
Intel C compiler 18.0.5,
Intel MPI version 2018 update 4 and
GSL 2.4.

If no protective measures are taken,
node failures can cause the loss of all the computation invested in the solution of a linear system.
However, they are a relatively rare occurrence.
With the incidence estimations of \cite{herault2015}
(mean time between failures of 9 hours for \mbox{100\,000} nodes,
and 53 minutes for \mbox{1\,000\,000} nodes),
a linear solver might be affected by at most a few of such events during its runtime.
Thus, we consider that examining the behavior of the solver
when a single node-failure event strikes at some point during its operation is a useful exercise.

We use a block Jacobi preconditioner,
with non-overlapping blocks and all rows of a block belonging to a single node.
The blocks are uniformly sized and we use as few of them as possible,
with a maximum block size of 10.
This preconditioner is used both for the linear system of the problem we are solving,
and for the inner systems of the reconstruction
(Lines \ref{alg:esr:pc_solve} and \ref{alg:esr:pcg:2:solve} of Alg \ref{alg:esr}).
The solver has converged once the relative residual $\|r\|_2 / \|b\|_2$ is below $10^{-8}$.
The relative residual for the inner system for the reconstruction must reach $10^{-14}$ for convergence.

Our test problems are SPD matrices from the SuiteSparse Matrix Collection~\cite{Davis2011a}
(see Table \ref{table:matrices}).
They were selected based on their size,
to allow for comparisons with related work in \cite{pachajoa2019esr}.
\begin{table}
\caption{Test matrices from \cite{Davis2011a}}
\label{table:matrices}
\begin{tabular}{llrr}
\toprule
\textbf{Matrix} & \textbf{Problem type} & \textbf{Problem size} & \multicolumn{1}{l}{ \textbf{\#NZ}} \\
\midrule
Emilia\_923 & Structural  &   923 136     & 40 373 538 \\
audikw\_1   & Structural  &   943 695     & 77 651 847 \\
\bottomrule
\end{tabular}
\end{table}

With these matrices, we set up the test constellation as follows:

\begin{itemize}
\item Two recovery strategies: ESRP and in-memory CR.
\item Checkpoint interval of 20, 50 and 100 iterations, plus an interval of 1 for ESRP,
representing the previously existing ESR method.
\item Resilience with 1, 3 and 8 redundant copies.
\item Reference runs, runs with resilience but without node failures,
and node failures introduced in contiguous blocks starting in ranks 0 and 64,
with as many node failures as the solver can tolerate with the number of available copies.
\end{itemize}

\noindent We introduce a node failure in the interval between checkpoints that contains the iteration $C/2$,
where $C$ is the number of iterations that a failure-free solver needs to converge.
Within this interval, the node failure is introduced two iterations before its end,
thus representing a worst-case scenario in which most of the progress since the start of the interval is lost.
Experiments are repeated at least five times for every setting in this test constellation.

The use of contiguous blocks of ranks for the node failures
is justified by considering that multiple-nodes failures would most likely come from, for example, a switch fault,
affecting a branch of the fat-tree and, consequently, a contiguous block of ranks.

\subsection*{Experimental results}

\begin{table*}[ht]
\caption{Results for matrix \texttt{Emilia\_923}. Reference time $t_0$ = 14.66 s.
The reference case takes $C$ = 10279 iterations to reach convergence.
The strategies shown are ESR with periodic storage (ESRP)
and In-memory buddy CR (IMCR).
$T$: Checkpointing interval, measured in iterations.
$\nredu{}$: Number of supported node failures.
$\nfail{}$: Number of introduced node failures.
All overheads are relative to $t_0$.
\emph{Failure-free overhead}: runtime overhead
of runs with resilience, but without introduced node failures.
The Location column indicates where the failures are introduced.
Rows marked with \emph{start} and \emph{center} have node failures introduced in blocks starting in ranks 0 and 64, respectively.
\emph{Overhead with node failures}: overall overheads for runs with an event
where as many nodes fail simultaneously as the solver can tolerate.
\emph{Reconstruction overhead}: overhead for the reconstruction operations
(collecting data in the replacement nodes and reconstructing the state for ESRP,
and sending the checkpointed data to the replacement node in IMCR).
All results are the median of at least five repeated experiments for the corresponding setting.
In all cases, node failures are introduced two iterations before a checkpoint for the interval containing the iteration \(C/2\),
thus representing a worst-case scenario.
The overhead for wasted iterations for \(T>1\) is not explicitly shown.
It can be approximated by subtracting the reconstruction overhead from the overall overhead since,
in general, the reconstruction does not change the trajectory of the solver after rollback.
}

\includegraphics[width=\textwidth]{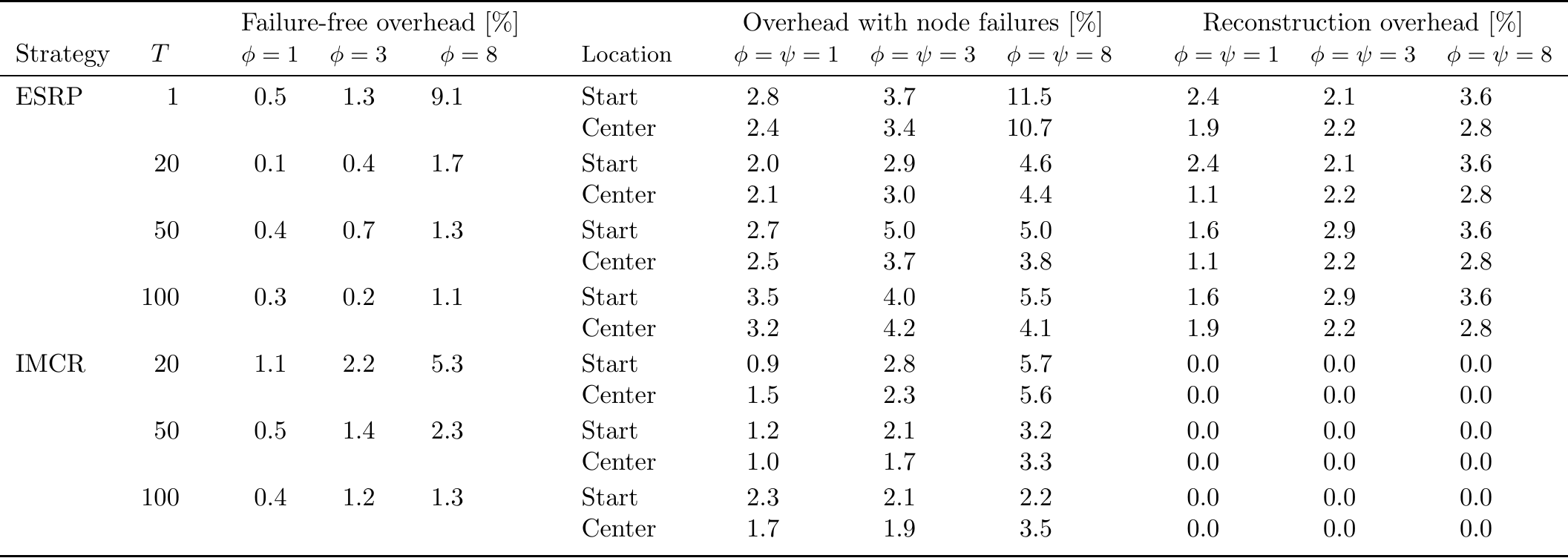}

\label{tab:emilia}
\end{table*}

\begin{table*}[ht]
\caption{Results for matrix \texttt{audikw\_1}. Reference time $t_0$ = 23.22 s.
The reference case takes $C$ = 5543 iterations to reach convergence.
Symbols and terms are explained in the caption of Table \ref{tab:emilia}
}

\includegraphics[width=\textwidth]{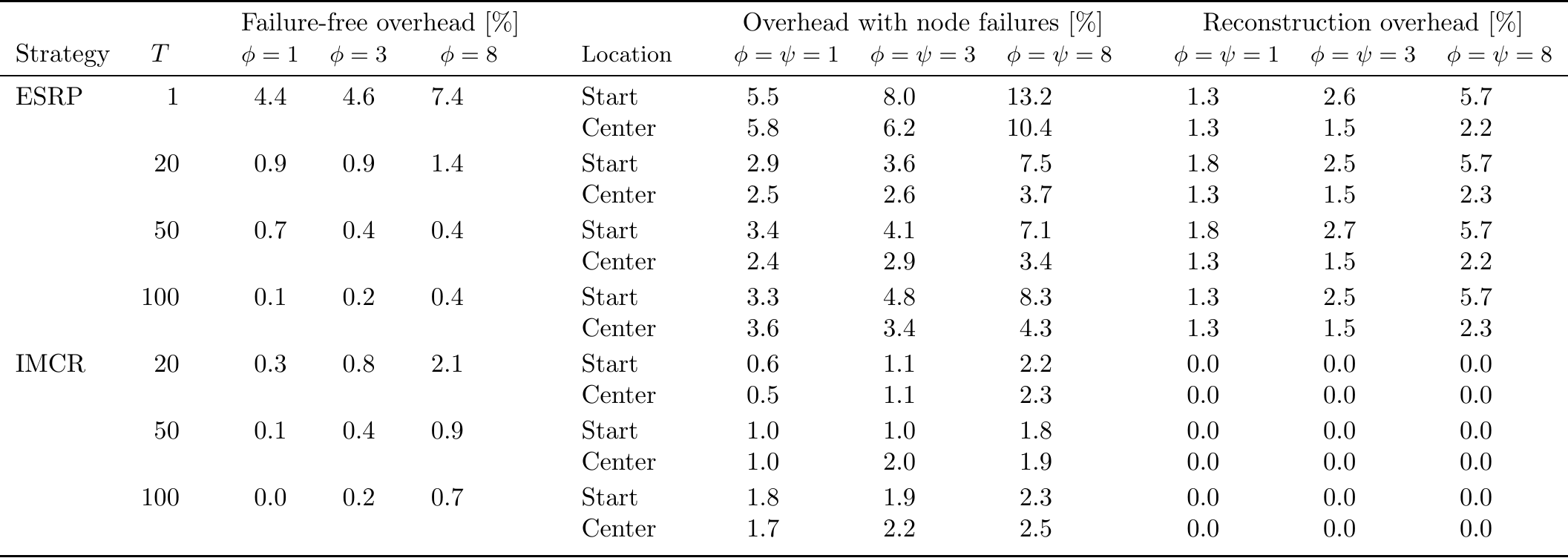}

\label{tab:audikw}
\end{table*}

\newcommand{\plotscale}{0.7}

\begin{figure*}[ht]
\centering
\includegraphics[width=\textwidth]{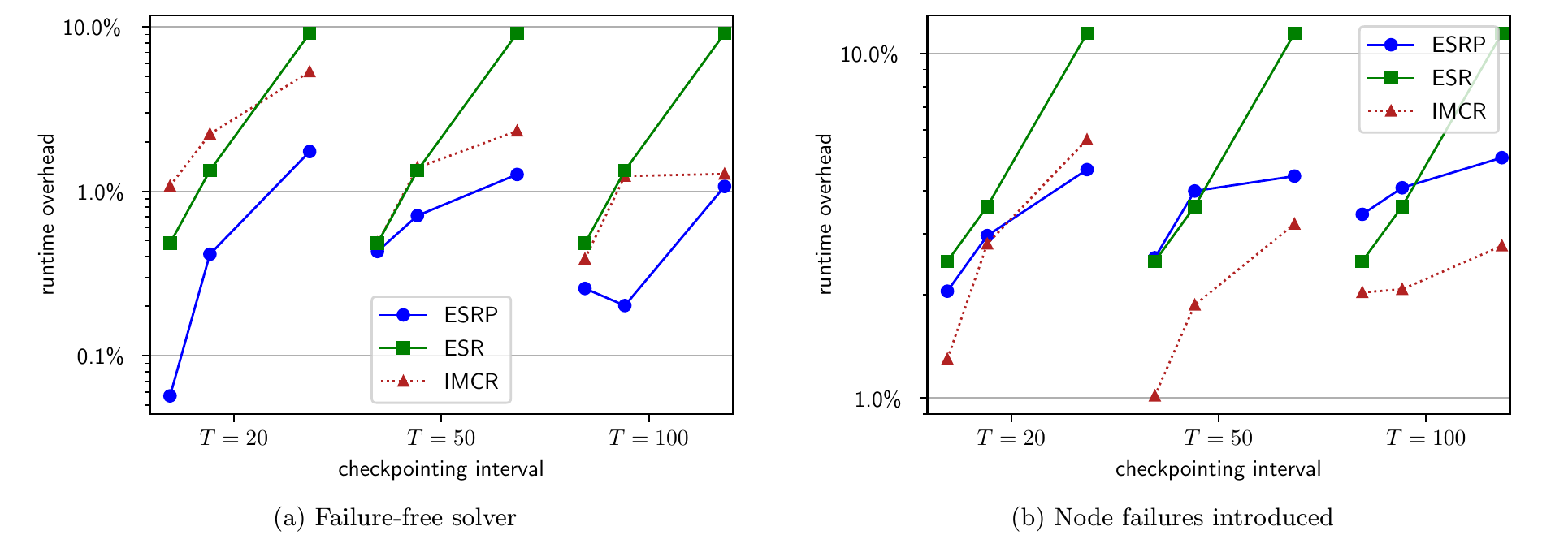}
\caption{Median runtime overhead for the experiments with the matrix \texttt{Emilia\_923}.
In each plot, experiments for a checkpoint interval $T$ are clustered together.
In each cluster, there are three lines,
representing experiments with ESRP, ESR and in-memory CR (IMCR).
In each line, the three markers, from left to right,
represent experiments with 1, 3, or 8 redundant copies,
and also 1, 3 or 8 simultaneous node failures
for Fig. \ref{fig:emilia}b, to the right.
ESR results are the same for all checkpointing intervals in each plot because they
are equivalent to ESRP results with $T=1$,
and are displayed along data for ESRP and IMCR for comparison.
The markers represent the median for results in all locations (cf. Tab~\ref{tab:emilia}) and repetitions.
}
\label{fig:emilia}
\end{figure*}

\begin{figure*}[ht]
\centering
\includegraphics[width=\textwidth]{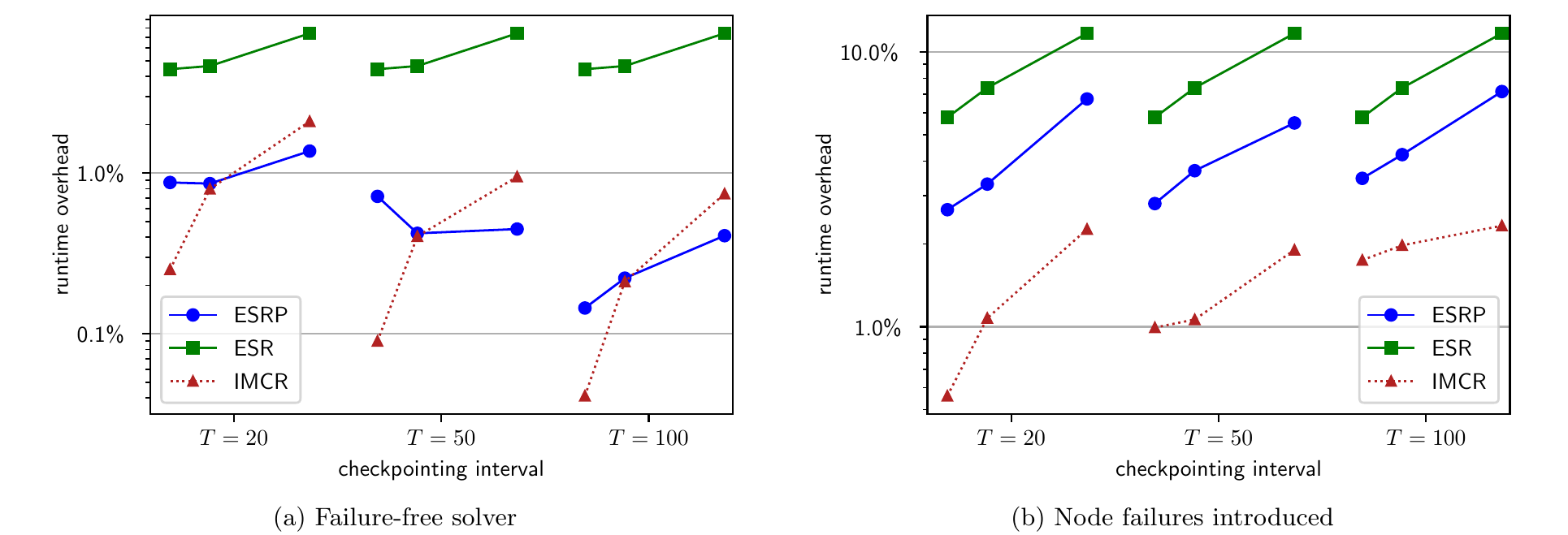}
\caption{Median runtime overhead for the experiments with the matrix \texttt{audikw\_1}.
See the caption of Fig. \ref{fig:emilia} for details on the structure of the plot.
}
\label{fig:audi}
\end{figure*}

We measure the runtime $t$ of the solver to reach convergence to evaluate our algorithms.
We define the reference time, \(t_0\), as the median time for runs of a reference, non-resilient PCG solver.
The metric that we present in our results is the relative overhead over this reference time:
\(\text{realtive overhead} = (t-t_0) / t_0\).
Our results are summarized in Tables \ref{tab:emilia} and \ref{tab:audikw},
and shown in Fig.~\ref{fig:emilia}~and~\ref{fig:audi}.
Note that in all of our experiments the measured total runtime is less than a
minute and, thus, well below an estimated mean time between failures of
approximately an hour up to several hours in large-scale applications
\cite{herault2015}.
Since we have to recover from a node failure in that short runtime, the
overhead due to this recovery is likely more severe than in a real-world
application with less frequent recoveries.
Therefore, we test a scenario that is \emph{less favorable} for our approach.
Reasonably low overheads in our experiments can hence be considered a proof of
concept.

For the used test matrices,
overheads of ESRP are usually much smaller than the ones of ESR (Section \ref{sec:background:esr}),
and this advantage is more pronounced for larger numbers of redundant copies.
In the case of ESRP, the columns for reconstruction overhead in Tables \ref{tab:emilia} and \ref{tab:audikw}
show the cost of gathering the information and of recomputing the lost data.
In the case of IMCR,
these columns show the cost of communicating checkpointed data to the replacement nodes.
For both test matrices,
our experiments show a reconstruction overhead of basically zero for IMCR.
This suggests that in our experimental setup
the communication cost is considerably smaller than the computation cost.
Keeping in mind the common understanding that communication tends to become more expensive than computation,
especially in the large scale,
this observation indicates that the experimental setup currently available to us
possibly leads to underestimating the communication cost
and does not allow for a solid comparison with IMCR.
Consequently, we decided to depict experimental measurements for IMCR
less prominently in Figs. \ref{fig:emilia} and \ref{fig:audi}
and leave more representative experiments at larger scales for future work.

The cluster that we use introduces a certain amount of variation to our measurements.
We repeat each experiment to reduce the standard deviation of the tests,
and we present their median.
However, there are cases in which this standard deviation for the reference runtimes
is not reduced below the overhead.
Table \ref{tab:emilia} shows an instance of this:
The median reference time for ESRP for three redundant copies
is larger for a period of 50 that for a period of 20.
In general, we expect larger checkpoint intervals to produce smaller overheads in the failure-free case,
but these results can be affected by the variation in runtime from external factors.
In this case, the overhead is so close to zero that it is overshadowed by the noise from the machine.

Table \ref{tab:emilia} shows that,
for the matrix \emilia{}, the failure-free overhead for ESRP is lower than for IMCR,
down to about half of the corresponding value for IMCR in some settings.
For ESRP, reducing the frequency at which redundant copies are stored visibly reduces the overhead,
especially in cases with multiple-nodes failures
(Tables \ref{tab:emilia} and \ref{tab:audikw}).
As for matrix \audi{}, Table \ref{tab:audikw} shows that overheads for the failure-free cases for ESRP and IMCR
are close, with some advantage for ESRP in cases with multiple-node failures.

In the case of ESRP, the ranks of the lost nodes determine the submatrix $\A{}_{\Is{\f{}},\Is{\f{}}}$
(where $\f{}$ represents the set of indices affected by the node failure),
and thus which inner linear system will be solved during the reconstruction
in Line \ref{alg:esr:pcg:2:solve} of Alg. \ref{alg:esr},
and how fast this can be done.
This cost is also influenced by the performance of the preconditioner used for the inner system.
As a consequence, the recovery times for ESRP change for different matrices
and for different sets of lost nodes for the same matrix.
In contrast, the recovery cost in IMCR is the cost of transferring checkpointed vectors
to the replacement nodes,
and it is more or less independent of the location of the failure.
Currently, our experiments use a simple block Jacobi preconditioner.
We believe that ESRP would greatly benefit from more appropriate preconditioners,
and an investigation of this aspect is future work.

Whether ESRP or IMCR is a better strategy for resilience depends on the probability
of node failures happening.
If the probability is low,
a method with a low overhead in the undisturbed case would be preferable,
even if the reconstruction cost is higher,
since it is unlikely that the runtime overhead must be incurred.
Conversely, a method with a lower recovery cost is preferable if the probability
of encountering a node failure is higher.

\subsubsection*{Accuracy of the experiments}
In general, when working with PCG without residual replacement, there is some drift between the vector \res{} and the vector $\B{}-\A{}\x{}$~\cite{vdvorst2000residual}.
In ESRP, solving the inner system with an iterative solver,
and performing the reconstruction in floating-point arithmetic,
cause the reconstructed vector \res{} not to be exactly equal to its state before the node failure.
To evaluate the accuracy of ESRP,
we compute the vector \(\B{}-\A{}\xs{End}\),
where the superscript \emph{End} represents the vectors' state after convergence,
and use it to compute the \emph{residual drift} metric:

\begin{equation}
\frac{\|\ress{End}\|_2 - \|\B{}-\A{}\xs{End}\|_2}{\|\B{}-\A{}\xs{End}\|_2}.
\end{equation}

A more positive value of the residual drift indicates a smaller value of \({\|\B{}-\A{}\xs{End}\|_2}\)
and, thus, a more accurate result.
This metric is not used to determine convergence of the solver;
for this, we use the relative residual as described in Section \ref{sec:experiments}.
The residual drift is computed only after the solver has converged.
We use this metric to ensure that ESRP is not generally less accurate than PCG.

PCG and ESRP experiments without node failures produce the same value for the metric
since they always follow exactly the same trajectory.
With node failures,
the residual drift depends on the selection of affected nodes and on the iteration when the node failure occurs;
in this case, we present the minimum and median for all experiments.
Accuracy results are summarized in Table. \ref{table:acc_summary}.
In the median, ESRP with node failures does not differ significantly from PCG.
As for the minimum value,
the results for the matrix \emilia{} show little accuracy loss with respect to PCG,
but for the matrix \audi{},
there is a drift of close to \(15.5\%\),
where PCG has a drift of close to \(8\%\).
We do not consider this to be a significant issue.
The slight advantage for ESRP in the median case for \audi is explained by the fact that
it reconstructs the iterand from the residual vector, thus making the residual and iterand consistent with each other.

\begin{table}[ht]
\caption{Residual drift observed in the experiments.
\emph{Reference}: Residual drift for all failure-free cases.
\emph{Median}: Median residual drift over all experiments with node failures.
\emph{Minimum}: Minimum residual drift over all experiments with node failures,
representing the greatest loss of accuracy during the reconstruction  in ESRP.
}

\begin{tabular}{llll}
\toprule
\multicolumn{1}{c}{Matrix} & \multicolumn{1}{c}{Reference} & \multicolumn{1}{c}{Median} & \multicolumn{1}{c}{Minimum} \\
\midrule
\emilia & \(-4.43\times 10^{-2}\) & \(-4.74\times 10^{-2}\) & \(-5.63 \times 10^{-2}\) \\
\audi   & \(-7.98\times 10^{-2}\) & \(-6.67\times 10^{-2}\) & \(-1.55 \times 10^{-1}\) \\
\bottomrule
\end{tabular}

\label{table:acc_summary}
\end{table}

\section{Conclusions and perspectives}
\label{sec:conclusions}

In this paper, we introduce an extension
to the exact state reconstruction (ESR) algorithm
for node-failure resilience for PCG.
Our approach reduces the runtime overhead of ESR by saving redundant copies
not in every iteration, but only every $T$ iterations.
Given the relation of this approach to checkpoint-restart (CR),
we call such a strategy an algorithm-based checkpoint-restart method.
We introduce a framework for our experiments
and evaluate this new strategy, which we call ESRP,
comparing it to standard ESR and also to our implementation of in-memory buddy CR (IMCR).
In our experimental results,
the runtime overhead of ESRP turns out considerably lower than that of standard ESR
and also lower than that of IMCR for the failure-free cases,
When node failures occur, however,
the recovery time is dominated by the solution of a smaller inner linear system
and depends on the matrix itself.

An important step to take in future work is to evaluate
ESRP using different preconditioners.
Furthermore, we are working on producing larger test problems,
so that we can observe a different regime in the computation/communication ratio for PCG.
Another interesting direction is the study of ESRP working with partitioning algorithms,
looking in particular for partitioning strategies that optimize for the matrix-vector product and simultaneously provide sufficient redundancy.
In the future, we intend to produce an implementation of the algorithms on a framework that can detect node failures and provide replacements,
using tools like ULFM~\cite{bland2013ulfm} and similar, for example.

\begin{acks}
This work has been funded by the \grantsponsor{wwtf}{Vienna Science and
Technology Fund (WWTF)}{https://www.wwtf.at/} through project
\grantnum{wwtf}{ICT15-113}.
The computational results presented have been achieved using resources
of the Vienna Scientific Cluster (VSC).
We thank our reviewers for their helpful comments and suggestions.
\end{acks}

\bibliographystyle{ACM-Reference-Format}
\balance
\bibliography{Bibliography.bib}

\end{document}